# All-in-one Spintronic Sensory Neurons for Magnetic Edge Intelligence

Jiajia Jian[1#], Bin Rong[2#], Jiaxin Liu[1], Fangyi Zhao[1], Kai Li[3], Tao Huang[1], Xue Zhou[1], Xin Yue[1], Zeyu Ma[1], Xuefeng Zheng[4], Weidong Zhang[5], Jian Fu Zhang[6], Yumeng Yang[2,*] and Zheng Chai[1,*]

[1] State Key Laboratory for Mechanical Behavior of Materials, and School of Materials Science and Engineering, Xi'an Jiaotong University, Xi'an, 710049, China.

[2] State Key Laboratory of Quantum Functional Materials, School of Information Science and Technology, ShanghaiTech University, Shanghai 201210, China

[3] School of Microelectronics and Physics, Hunan University of Technology and Business, Changsha, 410205, China.

[4] Faculty of Integrated Circuit, Xidian University, Xi'an, 710126, China.

[5] School of Engineering and Built Environment, Liverpool John Moores University, L3 3AF Liverpool, U.K.

[6] School of Integrated Circuits, Southeast University, Wuxi, 214026, China.

Jiajia Jian and Bin Rong contributed equally to this work.

Correspondence: Yumeng Yang (yangym1@shanghaitech.edu.cn) and Zheng Chai (zheng.chai@xjtu.edu.cn)

## Abstract

Magnetic fields enable non-contact sensing through nonmagnetic materials, supporting magnetic edge intelligence in autonomous recognition, navigation, and control. Magnetic edge intelligence requires efficient local sensing and processing, yet directly converting magnetic stimuli into neural spikes for neuromorphic computing remains challenging. Here, we demonstrate an all-in-one spintronic sensory neuron (SSN) based on spin-transfer torque magnetic tunnel junctions (STT-MTJs) fabricated in a 40-nm Magnetoresistive Random-Access Memory (MRAM) process. With magnetic-field-modulated stochastic switching, the SSN enables probabilistic sensing with a sensitive field window electrically shifted by write voltage, allowing adaptive operation across magnetic-field ranges without structural modification. This tunability alleviates the fixed sensitivity–range constraint of conventional tunneling magnetoresistance sensing, enabling a ±40 mT sensing range with an average sensitivity of 2.91%/mT. More importantly, such stochastic switching also directly transduces analog magnetic stimuli into neural spikes, integrating magnetic sensing, nonlinear rate encoding, and spike generation within the same device. The sigmoidal spike-rate response provides efficient magnetic-field encoding, verified by the neural engineering framework. With optimized encoding, a 10×10 SSN array achieves 90.7% accuracy for contactless magnetic trajectory recognition, approaching the software baseline. These results establish a scalable, MRAM-compatible route towards integrated sensing and neural processing for magnetic edge intelligence.

## 1. Introduction

Edge artificial intelligence (AI) increasingly requires physical stimuli to be sensed, converted, and processed locally under tight latency and energy constraints.[1,2] In conventional edge systems, sensing, conversion, and computation are usually carried out by separate hardware blocks, introducing substantial circuit complexity and energy overhead.[3] In contrast, biological systems follow a different strategy: sensory neurons directly transduce physical stimuli into neural spikes for subsequent neural processing. [4,5] Inspired by this principle, artificial sensory neurons have emerged in edge intelligence systems to directly encode physical stimuli into spike trains for downstream neuromorphic computing,[6,7] typically spiking neural networks (SNNs),[8] with demonstrations spanning optical,[9,10] pressure,[11] and thermal signals.[12–14]

Magnetic fields represent another important sensory modality because they enable non-contact detection and penetrate nonmagnetic materials. Magnetic sensing is therefore widely used in transportation,[15,16] industrial automation,[17,18] and biomedical applications, with emerging applications including unmanned aerial vehicle (UAV) navigation, robotic pipe inspection, and magnetically guided microrobots.[19] As these applications become increasingly autonomous, magnetic sensing must evolve beyond field detection to integrate sensing with local, task-oriented information processing. In this context, magnetic edge intelligence involves the sensing and processing of magnetic-field information at or near the sensing node to support tasks such as recognition, navigation, and control. A neuromorphic implementation requires two interconnected operations: directly converting magnetic stimuli into neural spikes and processing the resulting spikes for downstream computing. Integrating these operations offers a promising route toward compact, low-overhead magnetic sensing and intelligence at the edge.

Conventional Hall-effect and magnetoresistive sensors based on anisotropic magnetoresistance (AMR),[20] giant magnetoresistance (GMR),[21] or tunneling magnetoresistance (TMR)[22–24] generally produce analog electrical outputs that require analog-to-digital conversion before downstream spike-based neural processing (Figure 1a). Recent advances have explored low-power magnetic sensing using STT-MTJs[25,26] and spin-orbit torque (SOT)-driven nanomagnets,[27] as well as anomalous Hall effect devices integrating magnetic sensing and analog computing.[28] Although these approaches bring magnetic sensing and information processing closer together, direct transduction of magnetic stimuli into neural-compatible spikes through the sensing process itself and processing such device-native spike encoding for array-level neuromorphic computing remain largely unexplored. Establishing such a sensing-to-spike interface would allow magnetic information to enter downstream SNNs without separate multibit digitization and spike-generation stages, potentially reducing peripheral circuitry and system-level overhead.

Here, we introduce a stochastic magnetic sensing paradigm that exploits the thermally activated switching of STT-MTJs to directly encode magnetic stimuli into neural-compatible spikes. Under sub-threshold excitation, an STT-MTJ switches probabilistically between its magnetic states, a behavior previously exploited for Ising machines,[29] true random number generators (TRNGs),[30] and neuromorphic computing.[31] An external magnetic field modulates the switching energy barrier and hence the switching probability, allowing magnetic information to be represented by the statistics of stochastic switching events. Rather than extracting magnetic information from a resistance–field transfer characteristic, our approach encodes magnetic field in the switching probability. Electrical control of this probabilistic response enables the sensitive field window to be shifted across different field ranges without modifying the device structure or introducing feedback circuitry, thereby alleviating the fixed

sensitivity–range constraint of conventional TMR sensors. The sensing response covers a ±40 mT sensing range with an average sensitivity of 2.91%/mT.

More importantly, the same switching process also directly encodes analog magnetic stimuli into stochastic binary events. Based on this mechanism, we demonstrate an MRAM-compatible spintronic sensory neuron (SSN) fabricated using a 40-nm MRAM process (Figure 1b), in which repeated switching events generate spike trains whose rates intrinsically encode a magnetic field. The SSN thereby integrates magnetic sensing, nonlinear rate encoding, and spike generation within a single nanoscale memory device. Its physical encoding capability is experimentally validated by reconstructing magnetic inputs from measured spike trains using the Neural Engineering Framework (NEF). At the system level, a 10 × 10 SSN array modeled using experimentally measured device characteristics serves as the sensory front end of an SNN, which achieves 90.7% accuracy in contactless magnetic-trajectory recognition, approaching the performance of software-based reference models. By transforming the intrinsic stochasticity of STT-MTJs from switching uncertainty into a resource for sensory transduction and neural encoding, this work provides a scalable and manufacturing-compatible route toward sensor-native magnetic edge intelligence.

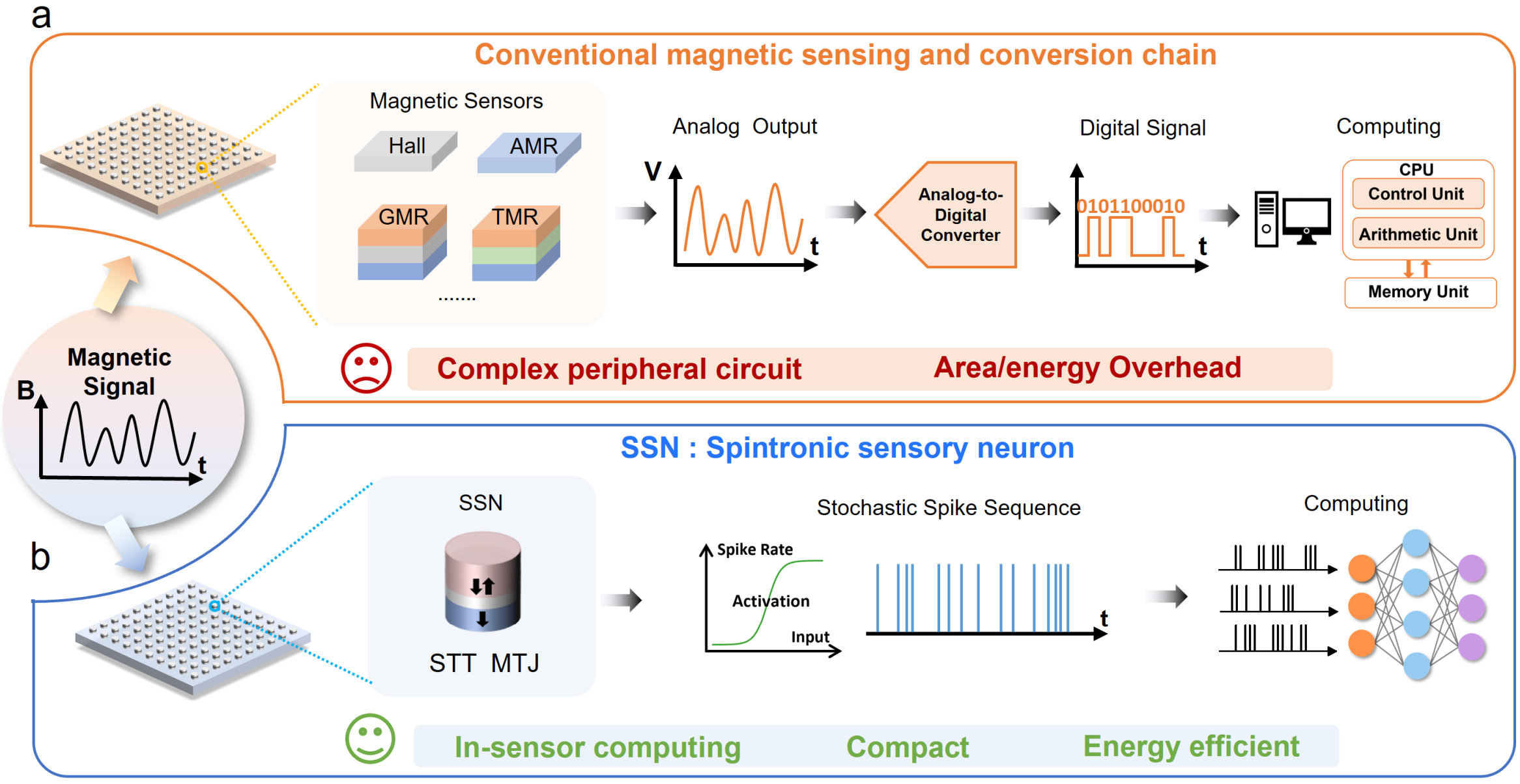

**Figure 1. Comparison of magnetic signal processing architectures.** (a) Conventional sensor–conversion–processor architecture, requiring separate signal conversion and neural processing circuits. (b) Proposed SSN based on standard STT-MTJs, integrating magnetic sensing, nonlinear activation, and stochastic spike encoding within a single device.

## 2. Results

## 2.1 Stochastic sensing scheme in STT-MTJs

The proposed stochastic sensing scheme is implemented using STT-MTJs fabricated on 12-inch wafers in a commercial foundry using a standard 40-nm MRAM process. Figure 2a shows a cross-sectional transmission electron microscopy (TEM) image of a 78-nm-diameter MTJ nanopillar, with corresponding EDX elemental maps provided in Supplementary Figure 1a. As illustrated in Figure 2b, the MTJ adopts a perpendicular CoFeB/MgO/CoFeB stack with a synthetic antiferromagnetic (SAF) reference structure. The MTJ exhibits two stable resistance states corresponding to the parallel (P) and antiparallel (AP) magnetization states (Figure 2c) and supports write pulses as short as 5 ns (Supplementary Figure 1).

Stochastic magnetic sensing originates from the dependence of STT stochastic switching on the magnetic field and write voltage. According to the Néel–Brown model,[32,33] for a write pulse of duration $t$, the switching probability of the MTJ is then given by

$$P_{sw} = 1 - \exp\left[-\frac{t}{\tau_0}\exp\left(-\frac{E_{P\to AP}(B,V)}{k_B T}\right)\right] \tag{1}$$

where $\tau_0$ is the attempt time, $k_B$ is the Boltzmann constant, $T$ is the temperature, and $E_{P\to AP}(B,V)$ is the effective switching barrier. Within a simplified thermally activated STT switching model, the effective barrier can be described as

$$E_{P\to AP}(B,V) = E_0\left(1-\frac{B}{B_K}\right)^2\left(1-\frac{V}{V_C}\right) \tag{2}$$

where $E_0$ is the zero-field energy barrier, $B$ is the externally applied magnetic field, $B_K$ is the effective anisotropy field, $V$ is the write-pulse voltage ($V_{write}$) and $V_C$ is the intrinsic critical voltage for P-to-AP switching. Increasing $V_{write}$ lowers the effective switching barrier through spin-transfer torque. An out-of-plane magnetic field $B$ is applied perpendicular to the MTJ plane, with the positive direction defined opposite to the initial P-state magnetization. A positive $B$ lowers the P-to-AP energy barrier, whereas a negative $B$ increases it (Figure 2d). The magnetic field and write voltage jointly modulate $P_{sw}$ through the effective switching barrier, providing the physical basis for stochastic magnetic sensing: rather than extracting the magnetic field from a continuous electrical response, the field is encoded directly into the probability of discrete MTJ switching events.

Experimentally, each reset–write–read cycle begins by resetting the MTJ to the P state, followed by a sub-threshold write pulse that induces a probabilistic P-to-AP transition and a subsequent state readout (Figure 2e). Repeating this cycle $N$ times at each magnetic field under fixed write-pulse yields the binary sequences as shown in Figure 2f. These sequences vary systematically with the applied magnetic field: positive fields ($B > 0$) promote P-to-AP switching and negative fields ($B < 0$) suppress it. The P and AP read-current levels are well separated, allowing individual switching events to be distinguished (Supplementary Figure 2). The field-dependent switching probability is calculated as $P_{sw} = \frac{N_{sw}}{N}$, where $N_{sw}$ is the number of successful P-to-AP switching events. Repeating the measurements over different magnetic fields $B$ and write voltages $V_{write}$ yields the two-dimensional $P_{sw}(B, V_{write})$ map shown in Figure 2g, with representative switching sequences shown in Supplementary Figure 3. This map provides a quantitative lookup table (LUT) relating $B$ and $V_{write}$ to $P_{sw}$.

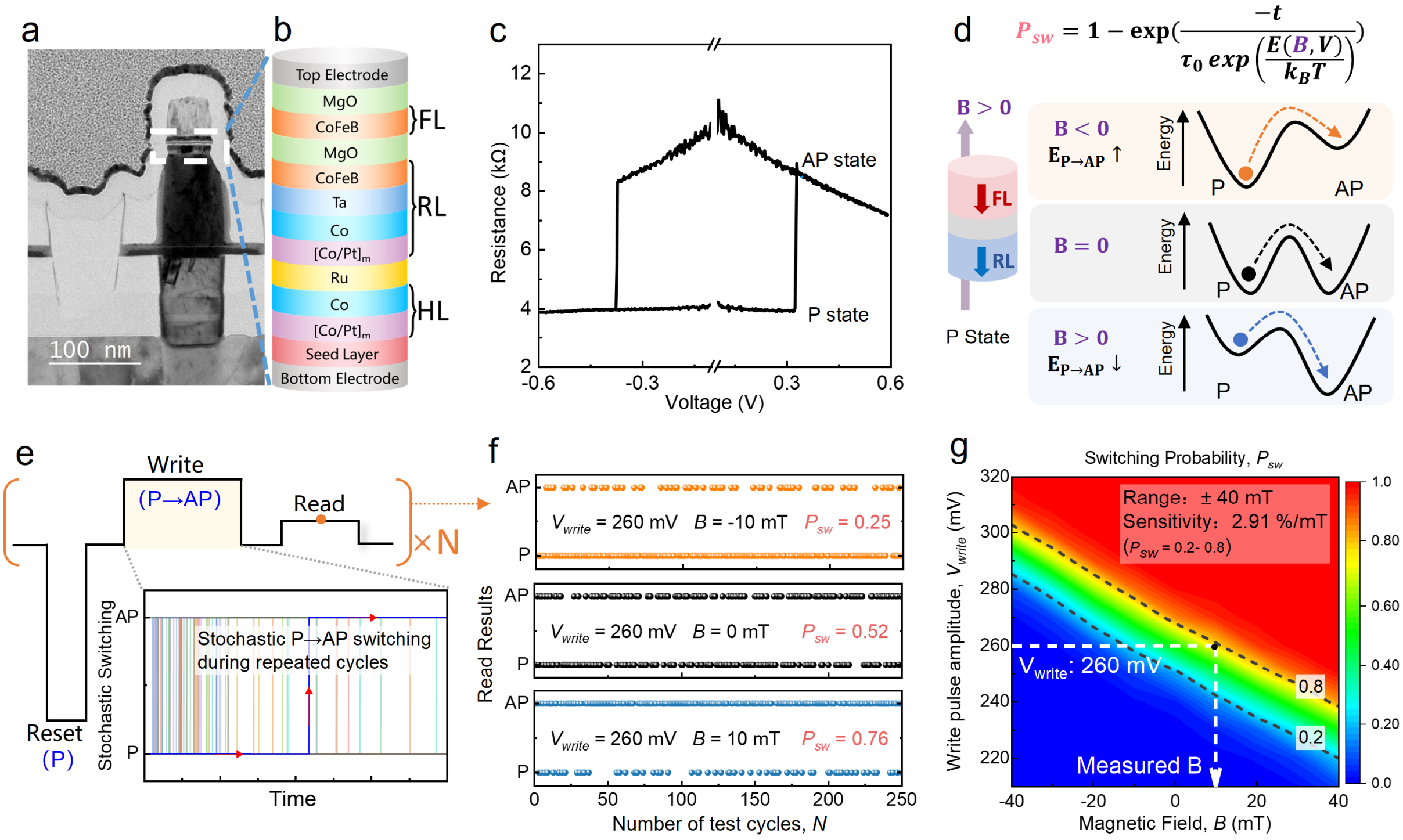

**Figure 2. MRAM-compatible STT-MTJ and magnetic-field-modulated stochastic switching.** (a) Cross-sectional transmission electron microscopy (TEM) image of a 78-nm-diameter MTJ nanopillar. (b) Schematic of the MTJ multilayer stack. (c) Measured resistance-voltage ($R$-$V$) hysteresis showing the stable parallel (P) and antiparallel (AP) resistance states. (d) Definition of the applied out-of-plane magnetic field $B$ and schematic energy landscapes illustrating the field modulation of the P-to-AP switching barrier. Positive fields ($B > 0$) lower the barrier, whereas negative fields ($B < 0$) increase it. (e) Stochastic P-to-AP switching under repeated reset–write–read cycles with sub-threshold write pulses. (f) Stochastic switching of the MTJ measured under different magnetic fields at $V_{write}$ = 260 mV. (g) Two-dimensional switching-probability map $P_{sw}(B, V_{write})$.

The $P_{sw}(B, V_{write})$ LUT covers a sensing range of $\pm$40 mT, with an average sensitivity = 2.91%/mT within the $P_{sw}$= 0.2−0.8 operating region. Here, the sensitivity is defined as $\frac{dP_{sw}}{dB}$ at a fixed write voltage. The LUT offers a favorable combination of sensing range and sensitivity compared with conventional magnetoresistive sensors (Table S1). These sensors generally face a trade-off between sensitivity and sensing range. Approaches based on magnetic biasing or closed-loop compensation can extend the sensing range by shifting the sensitive window, but require additional field-generating structures or feedback circuitry.[34,35] In our device, the sensing window is instead shifted directly by $V_{write}$. As $V_{write}$ varies, the steep region of the $P_{sw} - B$ response shifts along the magnetic-field axis, placing regions with a relatively large sensitivity at different field ranges. More importantly, the magnetic field is encoded in the statistics of these stochastic binary switching events, while each switching event can be directly used as a spike without multibit analog-to-digital conversion.

## 2.2 Stochastic Magnetic Sensing

The field dependence of $P_{sw}$ characterized in Figure 2 provides the basis for stochastic magnetic-field sensing. In each reset–write–read cycle, the MTJ state is read by a pre-charge sense amplifier (PCSA), with 0 representing the P state and 1 the AP state (Figure 3a). Repeated cycles generate a binary switching sequence (Figure 3b), from which $P_{sw}$ is calculated. The magnetic field $B$ is then extracted from $P_{sw}$ using the LUT established in Figure 2g. As the number of sensing cycles $N$ increases, the extracted magnetic field converges toward the applied value (Figure 3c). For an applied field of 10 mT, consistent readout is obtained by using $10^3$, $10^4$, and $10^5$ sensing cycles, without observable drift (Figure 3d).

Increasing $N$ improves the field resolution by reducing the statistical variation in $P_{sw}$. For $N$ independent stochastic switching events of MTJ, the standard deviation of the measured switching probability follows $\sigma_p \propto N^{-1/2}$.[36] The minimum detectable field is then estimated as[37]

$$B_{min} = \frac{\sqrt{S_P(f)}}{|dP_{sw}/dB|} \times \sqrt{\frac{f_{cycle}}{2N}} \qquad (3)$$

where $S_P(f)$ is the noise power spectral density of the binary switching sequence (Supplementary Figure 4) and $f_{cycle}$ is the reset–write–read cycle frequency. As shown in Figure 3e, increasing $N$ therefore improves the field resolution at the cost of a longer sensing time, providing a direct trade-off between sensing speed and resolution. Compared with representative magnetic sensors, this stochastic sensing approach combines an MRAM-compatible nanoscale STT-MTJ, nanosecond switching capability, and a broad magnetic-field range (Table S1).

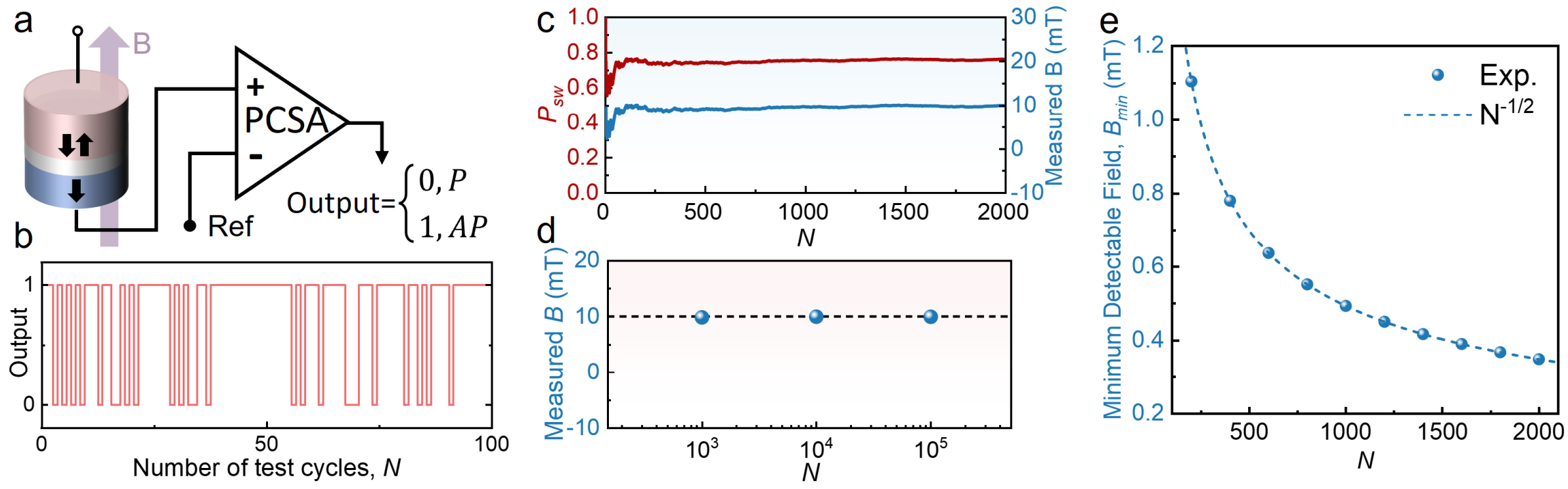

**Figure 3. Stochastic magnetic sensing using the STT-MTJ.** (a) Schematic of magnetic-field sensing through repeated reset–write–read cycles with PCSA readout. (b) Representative measured binary switching sequence. (c) Convergence of $P_{sw}$ and the extracted magnetic field with increasing number of sensing cycles $N$, validating statistical magnetic-field readout. (d) Cycling stability of stochastic magnetic sensing, showing drift-free 10 mT field readout with up to $10^5$ sensing cycles. (e) Minimum detectable field $B_{min}$ as a function of the number of sensing cycles $N$.

## 2.3 Neural Encoding with Stochastic Sensing Neurons

Beyond magnetic-field sensing, the stochastic switching events of the MTJ can be directly used as spikes without first calculating $P_{sw}$ or extracting the magnetic field. This direct conversion of magnetic inputs into stochastic spikes enables the MTJ to operate as an SSN (Figure 4a). At a fixed cycle frequency $f_{cycle}$, the spike rate is given by $r_{spike} = f_{cycle} \times P_{sw}$, corresponding to a commonly used rate-encoding scheme in neural networks.[38] Because $P_{sw}$ varies sigmoidally with the magnetic field, the spike rate also exhibits a sigmoidal response, providing an intrinsic nonlinear encoding function for downstream neuromorphic computing.[39] To evaluate how well the stochastic spike sequence represents the magnetic input, the SSN-generated spike sequence is analyzed using the Neural Engineering Framework (NEF).[40] The NEF decoding is used here only to evaluate the information encoded in the spike sequence and is not required for SNN processing. In the NEF decoding process, the generated spike sequence is convolved with an exponentially decaying synaptic kernel and summed to obtain a decoded magnetic-field signal $B_{decode}$ (Figure 4a).[41] The decoded signal $B_{decode}$ is then compared with the applied magnetic field $B_{in}$ to evaluate how accurately the magnetic input is represented by the spike sequence. Details of the decoding procedure are provided in Section 5 of Supporting Information.

We first apply constant magnetic fields and measure the corresponding SSN spike sequences at a write-pulse amplitude of 260 mV. NEF decoding is implemented in MATLAB. As shown in Figure 4b, the NEF-decoding output $B_{decode}$ agrees well with the input magnetic fields $B_{in}$, with a mean absolute error of 0.544 mT (1.36% FS) over the tested range from −20 to 20 mT. We next examine how the SSN spike sequence represents a sinusoidally dynamic magnetic field. Two strategies are evaluated to reduce the deviation between the input $B_{in}$ and output $B_{decode}$ (Figure 4c). This deviation is quantified using the normalized SSN encoding error. First, the SSN encoding ratio denotes the number of reset–write–read cycles used to encode the magnetic input within each time window. Increasing this ratio from 5 to 50 reduces the encoding error by approximately 50% (Figures 4d and 4e). Second, the stochastic outputs of multiple parallel MTJs are combined to form the output of a single SSN. Increasing the number of MTJs per SSN from one to five reduces the encoding error by approximately 50% (Figures 4f and 4g). The NEF results verify that the SSN can serve as a compact stochastic-spike front end and provide a direct path toward downstream SNN processing.

Owing to the binary MTJ output and minimal peripheral circuitry, the SSN eliminates the need for separate multibit digitization and stochastic spike-generation stages. The estimated transistor count is reduced by a factor of approximately 170 compared with a conventional digital sensor-to-spike chain (Supplementary Figure 5). By simplifying the conventional multistage sensing-to-spike chain, the SSN substantially reduces the circuit complexity of the sensing interface. This reduction becomes particularly important when scaling to large sensor arrays, where the peripheral circuitry associated with each sensing channel could introduce considerable hardware overhead.

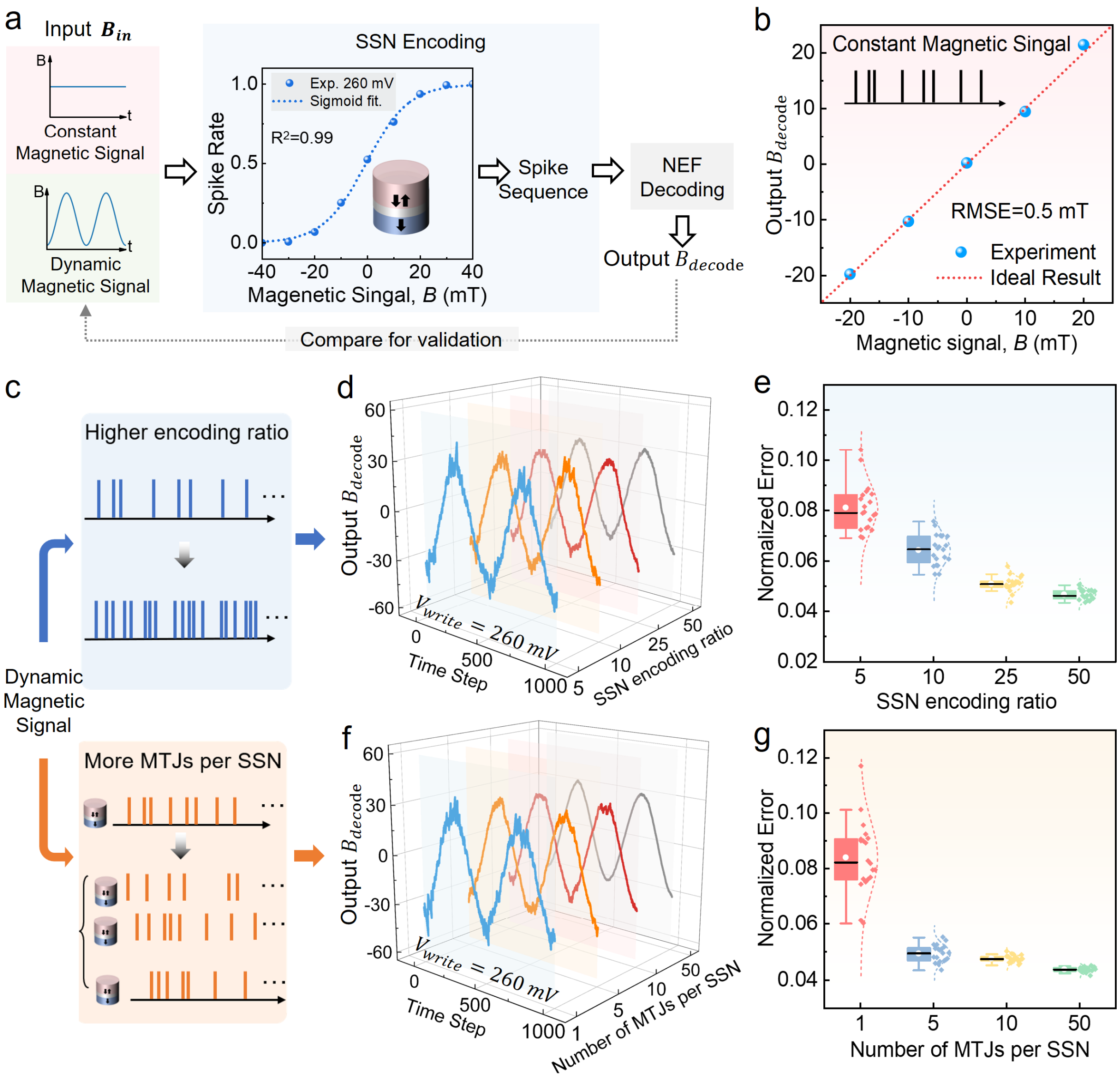

**Figure 4. Neural encoding of the SSN evaluated using the Neural Engineering Framework (NEF).** (a) Schematic of SSN encoding and NEF decoding. The SSN converts the magnetic input $B_{in}$ into stochastic spike sequence through rate encoding. The spike sequence is decoded by convolution and summation to obtain $B_{decode}$ for comparison with $B_{in}$. (b) Decoding of constant magnetic-field inputs from experimentally measured SSN spike sequences. (c) Two strategies for reducing the SSN encoding error for dynamic magnetic inputs: increasing the SSN encoding ratio and increasing the number of MTJs per SSN. (d, e) Decoded magnetic fields $B_{decode}$ and corresponding normalized SSN encoding errors at different SSN encoding ratios, with the number of MTJs fixed at one per SSN. (f, g) Decoded magnetic fields $B_{decode}$ and corresponding normalized errors for different numbers of MTJs per SSN, with the SSN encoding ratio fixed at 5.

## 2.4 SSN-Enabled SNN for Magnetic-Trajectory Recognition

To evaluate the potential of the SSN for neuromorphic computing, we simulate an SSN-enabled SNN for magnetic-trajectory recognition using experimentally measured SSN responses. The magnetic-trajectory dataset is generated using the UNIPEN handwriting database[42] by treating handwritten strokes as the motion paths of a simulated magnetic source. As the source moves above a simulated 10×10 SSN array, the magnetic field at each SSN is calculated using the simplified spatial magnetic-field profile,[43,44] producing a time-varying magnetic-field distribution across the array (Figures 5a and 5b). Each SSN converts its local magnetic-field input into a stochastic spike train, and the resulting 100 spike sequences are fed directly into the hidden layer of SNN for trajectory classification (Figures 5c and 5d).

Figures 5e and 5f show the confusion matrices after 5 and 30 training epochs, respectively. With an encoding ratio of 5 and one MTJ per SSN, the SNN achieves a test accuracy of 87.2% after 60 epochs (Figure 5g). We further optimized SSN encoding by increasing the ratio to 20 and using two MTJs per SSN, improving the test accuracy to 90.7%. Compared with the original encoding configuration, the optimized SNN also reaches the same accuracy with approximately 53% fewer training epochs. For comparison, a direct-encoding software SNN achieves a test accuracy of 92.1% on the same task. These results show that the SSN-generated spikes can support SNN-based processing with classification performance comparable to software-based reference models.

By exploiting magnetic-field-modulated stochastic switching, the SSN combines magnetic sensing, nonlinear encoding, and spike generation within the same STT-MTJ. Importantly, as the building blocks of STT-MRAM, STT-MTJs have benefited from extensive technological development with demonstrated scalability[45,46] and CMOS compatibility,[47–49] as well as nanoscale dimensions, nanosecond switching speeds, and high endurance.[50] These advantages provide a practical foundation for scaling SSNs into larger arrays. In the future, such arrays could be further integrated with CMOS logic or MTJ-based in-memory computing circuits to combine sensing and downstream processing on the same hardware platform.

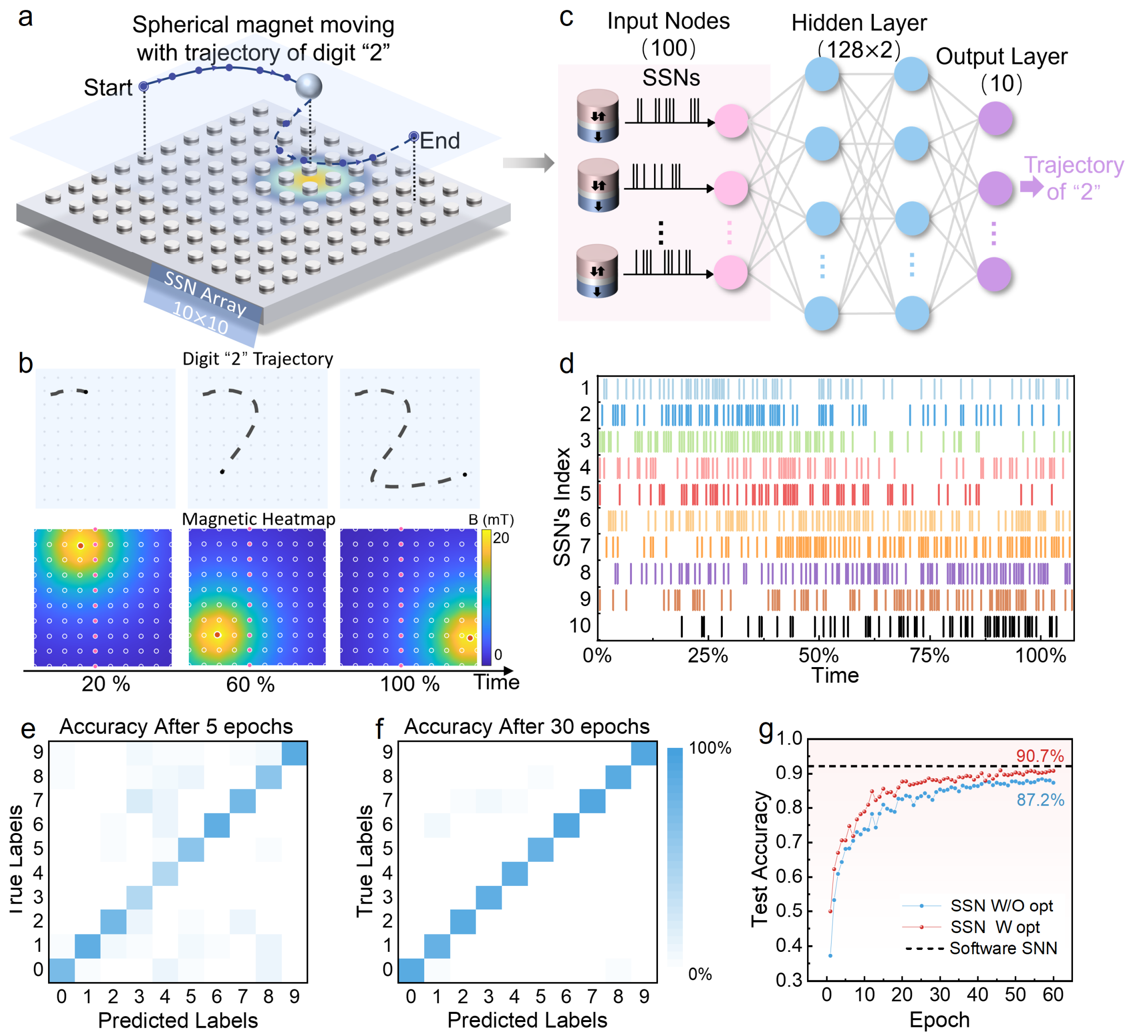

**Figure 5. Magnetic-trajectory recognition using the SSN-based SNN.** (a) Schematic of the magnetic-trajectory recognition task using a simulated 10 × 10 SSN array. (b) Trajectory snapshots and time-varying magnetic-field distribution across the SSN array. (c) SNN architecture using SSN-generated spike sequences as the input for trajectory classification. (d) Representative SSN spike sequences generated at $V_{write}$ = 250 mV. (e,f) Confusion matrices after 5 and 30 training epochs with encoding optimization, respectively. (g) Test accuracy of the SSN-based SNN (without and with encoding optimization) and the software SNN reference.

## 2.5 Conclusion

We demonstrate an SSN based on STT-MTJs fabricated using a 40-nm MRAM process. Unlike conventional TMR sensing based on a fixed resistance–field response, the proposed stochastic sensing paradigm encodes magnetic fields in switching probability, enabling an electrically tunable sensitive window across different field ranges. The same switching process enables the device to perform adaptive magnetic sensing, nonlinear rate encoding, and binary spike generation, directly bridging magnetic stimuli and downstream spike-based neuromorphic computing. This unified sensing-and-encoding mechanism eliminates separate multibit digitization and spike-generation stages, substantially simplifying the sensing-to-spike peripheral circuitry. System-level validation further demonstrates accurate contactless magnetic-trajectory recognition with strong robustness to device variation. These results establish stochastic switching as a new paradigm for programmable magnetic sensing and neural encoding, and standard STT-MRAM as a scalable, manufacturing-ready platform for sensor-native edge intelligence.

## 2.6 Experimental Section

### Device Fabrication

The perpendicular STT-MTJs incorporated a Co/Pt-based synthetic antiferromagnetic (SAF) structure, a CoFeB reference layer, an MgO tunnel barrier, and a CoFeB free layer. The multilayer stack was deposited on a NiCr seed layer and subsequently annealed in vacuum. The films were then patterned into circular MTJ nanopillars using standard lithography and etching processes. Cross-sectional transmission electron microscopy (TEM) and energy-dispersive X-ray spectroscopy (EDX) were used to characterize the device structure and elemental distributions.

### Electrical Characterization under Magnetic Fields

Electrical characterization was performed using a Keithley 4200-SCS semiconductor parameter analyzer equipped with pulse measurement units (PMUs). The resistance–voltage (R–V) characteristics were obtained using pulsed I–V measurements, in which write pulses with swept amplitudes were interleaved with low-voltage read pulses to determine the MTJ resistance state. Stochastic switching was characterized using repeated reset–write–read cycles. In each cycle, the MTJ was first reset to the P state, then subjected to a sub-threshold write pulse, and finally read to determine its resistance state. Unless otherwise specified, the write-pulse width was fixed at 500 ns, and each reset–write–read cycle had a total duration of 1.5 μs, corresponding to a cycle frequency ($f_{cycle}$) of approximately 0.667 MHz.

A controlled magnetic-field environment was provided by a CPX-VF superconducting-magnet probe station. The magnetic field was applied perpendicular to the MTJ plane, with the positive direction defined as shown in Figure 2d. The electrical pulse sequence and magnetic-field control were automated through LabVIEW. All measurements were performed at room temperature.

## Magnetic-Trajectory Dataset Construction

The magnetic-trajectory dataset was constructed from the UNIPEN online handwriting database, a standardized dataset developed for handwriting-recognition research.[42] Unlike image-based handwriting datasets, UNIPEN records the writing process as temporally ordered pen trajectories, including sequences of two-dimensional pen coordinates and associated pen-motion information. The isolated-digit category, consisting of handwritten trajectories for digits 0–9, was used in this work. The 7,274 trajectories were randomly divided into 5,819 training samples and 1,455 test samples. These trajectory data preserve the spatial path and temporal order of the original handwriting and were used to define the motion of a simulated magnetic source above the 10×10 SSN array.

Each trajectory was spatially normalized and mapped onto a 10×10 SSN array with an SSN pitch of 1 mm. The magnetic field of the moving source was modeled using a simplified spatial profile,[43,44] with a maximum out-of-plane magnetic field of $B_{max} = 20$ mT directly beneath the source. At each time step, the out-of-plane magnetic field applied to the $i$-th SSN was calculated as

$$B_{z,i}(t) = B_{max} \frac{h^3}{(d_i^2(t) + h^2)^{3/2}} \tag{4}$$

where $h$ is the vertical distance between the magnetic source and the SSN array and $d_i^2(t)$ is given by

$$d_i^2(t) = (x_i - x_m(t))^2 + (y_i - y_m(t))^2 \tag{5}$$

denotes the squared in-plane distance between the $i$-th SSN at$(x_i, y_i)$ and the instantaneous position of the magnetic source $(x_m(t), y_m(t))$.

## Spiking Neural Network Simulations

The resulting time-varying magnetic fields were converted into stochastic spike sequence using the experimentally measured field-dependent SSN response. The spike sequence generated by the 10×10 SSN array was directly fed into the SNN and propagated through two hidden layers followed by an output layer. For comparison, a direct-encoding SNN with the same architecture as the SSN-based SNN was implemented as a baseline model, comprising 100 input neurons, two hidden layers with 128 neurons each, and 10 output neurons.

The magnetic-trajectory preprocessing and magnetic-field calculation were implemented in MATLAB. The SNN was implemented in Python using PyTorch and the SpikingJelly framework, with GPU acceleration supported by CUDA. Data processing and performance evaluation were performed using NumPy, SciPy, and scikit-learn.

## Acknowledgements

This work is financially supported by the National Key R&D Program of China (Grant No. 2025YFA1016400), the National Natural Science Foundation of China (Grant No. 12327806 and 62674189), Shanghai 2025 "Science and Technology Innovation Action Plan" Fundamental Research Program in Integrated Circuit Grant 25JD1403200.

## Author Contributions

Z.C. and Y.M.Y. designed the study. Z.C. supervised the project. J.J.J. performed the device measurement and experiments, conducted the SNN simulations, collected and analyzed the data. J.X.L. contributed to the experiments. B.R. contributed to data analysis. K.L. contributed to the circuit analysis. J.J.J., B.R., Z.C., and Y.M.Y. prepared, reviewed, and revised the manuscript. All authors contributed to the discussion, provided feedback on the manuscript, and approved the final version.

## Data Availability Statement

The data that support the findings of this study are available from the corresponding author upon reasonable request.